\RequirePackage{fix-cm}
\documentclass[smallextended]{svjour3}       
\smartqed  
\usepackage{graphicx}
\begin{document}

\title{Variable coefficient complex Ginzburg-Landau equation
}


\author{Yusuke Uchiyama}


\institute{1-60-20 Minami-Otsuka, Toshima-ku, Tokyo, Japan  \\
              \email{uchiyama@mazin.tech}           
}

\date{Received: date / Accepted: date}

\maketitle

\begin{abstract}
The complex Ginzburg-Landau equation (CGLE) is a general model of spatially extended nonequilibrium systems.  In this paper, an analytical method for a variable coefficient CGLE is presented to obtain exact solutions.  Variable transformations for space and time variables with coefficient functions yield an imaginary time advection equation related to a complex valued characteristic curve.  The variable coefficient CGLE is transformed into the nonlinear Schr{"\o}dinger equation (NLSE) on the complex valued characteristic curve.  This result indicates that the analytical solutions of the NLSE generate that of the variable coefficient CGLE.  
\end{abstract}
\section{Introduction}
\label{sec:1}
The complex Ginzburg-Landau equation (CGLE) is a general model of spatially extended nonlinear dissipative systems~\cite{Cross1993,Aranson2002}, such as, fluid convections~\cite{Daniels2004}, fiber optics~\cite{Akhmediev2005}, chemical reactions~\cite{Beta2006} and biological systems~\cite{Bekki2012}.  Near the principal Hopf-bifurcation point of the governing equation, the method of multiple scales provides the CGLE as a slowly varying amplitude equation, which describes both amplitude and phase dynamics on a carrier wave~\cite{Kuramoto1984}.  Thus the solutions of the CGLE have been studied to understand the nature of nonlinear dissipative systems. \par
The analytical methods for solving nonlinear wave equations have been developed, especially for the nonlinear Schr{\"o}dinger equation (NLSE), which is obtained as the non-dissipation limit of the CGLE~\cite{Aranson2002}.  Since the NLSE is a class of integrable systems, sophisticated analytical methods give the exact solutions: the inverse scattering method serves as a prescription for the initial value problem of the NLSE~\cite{Ablowitz1974}; the Hirota's bilinear method introduces algebraic procedures to obtian the special solutions of the NLSE~\cite{Hirota1971}.  Although the CGLE is not an integrable system, Bekki and Nozaki derived analytical solutions by developing a¡ modification of the Hirota's bilinear method~\cite{Bekki1984}.  \par
Related to the CGLE and NLSE, nonlinear wave equations with variable coefficients have attracted the attention of researchers recently.  In the area of the Bose-Einstein condensation, the dynamics of macroscopic wave functions are described by the NLSE with a potential function, which is known as the Gross-Pitaevskii equation (GPE)~\cite{Pethick2001}.  For specific forms of the potential functions the Hirota's bilinear method provides special solutions of the GPE~\cite{Li2007}.  A closed form of the parameter functions of the elliptic function is introduced to obtain a periodic function of the GPE~\cite{Bronski1,Mallory2013}.  The amplitude modulation of plasma systems, where an electron beam is injected under a high-frequency electric field, is desciribed by the NLSE with time-dependent coefficients~\cite{Gao2001}.  A similar variable coefficient NLSE was presented as a model of impulse transmission in inhomogeneous optical fiber~\cite{Liu2009}.  These variable coefficient NLSEs were solved in terms of sets of ordinary differential equations induced from appropriate ansatz.  As is often used as a general model of dissipative signal propagations in optical fiber, the CGLE with time-dependent coefficient functions has been introduced to describe laser dynamics in optical fiber~\cite{Fang2006,Wong2015,CHOW2008,Liu2017}.  The modified Hirota's bilinear method with a family of time-dependent functions has been introduced to obtain the analytical solutions of the CGLE with time-dependent coefficients.  \par
This paper presents an analytical method to solve the CGLE with space- and time-dependent variable coefficients by using the method of change of variables.  From this method, an imaginary time advection equation is derived and provides a characteristic curve on which the variable coefficient CGLE is transformed into the NLSE.  Solvable conditions of the variable coefficients CGLE are introduced, related to the imaginary time advection equation.  An example of transformed coefficients is demonstrated and related physical systems are discussed.  Finally the conclusion of this paper is provided.
\section{Formal solutions of the variable coefficient complex Ginzburg-Landau equation}
The CGLE is derived from nonlinear partial differential equations governing the dynamics of dissipative systems, based on the assumption that the spatiotemporal dynamics can be described by slow amplitude dynamics on carrier waves.  In order to incorporate the effects of inhomogeneity in media and temporal modulation, space- and time-dependent variable coefficients are introduced into the CGLE in the form:
\begin{equation}
\mathrm{i}\frac{{\partial}{\psi}}{{\partial}t} + p(x,t)\frac{{\partial}^2{\psi}}{{\partial}x^2} + q(x,t)|{\psi}|^2{\psi} = [-{\omega}(t) + \mathrm{i}{\gamma}(t)]{\psi},
\label{eq:VCGLE}
\end{equation}
where ${\psi}(x,t)$ is a complex valued field function of the space $x$ and time $t$ variables.  The coefficient functions $p(x,t)$ and $q(x,t)$ are given by $p(x,t)=p_r(x,t)+\mathrm{i}p_i(x,t)$ and $q(x,t)=q_r(x,t)+\mathrm{i}q_i(x,t)$, where $p_r(x,t)$, $p_i(x,t)$, $q_r(x,t)$, $q_i(x,t)$, ${\gamma}(t)$, and ${\omega}(t)$ are real valued functions.  It is assumed that $p(-x,t)q(-x,t)=p(x,t)q(x,t)$.  In the literature of nonlinear wave theory, $p(x,t)$, $q(x,t)$, ${\gamma}(t)$, and ${\omega}(t)$ correspond to linear dispersion and dissipation, nonlinear saturation, linear gain or loss, and frequency modulation coefficients, respectively.    \par
Hereafter, the analytical procedure of deriving formal solutions of the variable coefficient CGLE is demonstrated by means of variable transformations and the method of characteristics.
\subsection{Variable transformations}
\label{subsec:2-1}
To begin, the complex valued field function ${\psi}(x,t)$ is transformed into
\begin{equation}
{\psi}(x,t)={\exp}[{\Gamma}(t)+\mathrm{i}{\Omega}(t)]{\varphi}(x,t),
\label{eq:PSItoPHI}
\end{equation}
where ${\Gamma}(t)$ and ${\Omega}(t)$ are defined by
\begin{eqnarray}
&&{\Gamma}(t)=\int_{}^{t}{\gamma}(t')\mathrm{d}t',
\label{eq:GAMMA} \\
&&{\Omega}(t)=\int_{}^{t}{\omega}(t')\mathrm{d}t'.
\label{eq:OMEGA}
\end{eqnarray}
Substituting Eqs.~(\ref{eq:PSItoPHI}), (\ref{eq:GAMMA}) and (\ref{eq:OMEGA}) into Eq.~(\ref{eq:VCGLE}) with a transformed variable
\begin{equation}
{\tau}=\int_{}^{t}q(x,t')\mathrm{e}^{2{\Gamma}(t')}\mathrm{d}t'
\label{eq:TFTAU}
\end{equation}
and a transformed coefficient function
\begin{equation}
r(x,t)^2=\frac{p(x,t)}{q(x,t)}\mathrm{e}^{-2{\Gamma}(t)},
\label{eq:TFR}
\end{equation}
a variable coefficient NLSE is obtained as
\begin{equation}
\frac{{\partial}{\varphi}}{{\partial}{\tau}}+r(x,{\tau})^2\frac{{\partial}^2{\varphi}}{{\partial}x^2}+|{\varphi}|^2{\varphi}=0,
\label{eq:VCNLSE}
\end{equation}
where $r(x,{\tau})$ is the expression of $r(x,t)$ with respect to ${\tau}$.  If the $x$-derivative operator with the coefficient function $r(x,{\tau})$ is transformed into that with a constant parameter, Eq.~(\ref{eq:VCNLSE}) is reduced to the integrable NLSE.  \par
In order to obtain variable transformations which reduce Eq.~(\ref{eq:VCNLSE}) to the integrable NLSE, a complex valued characteristic curve is introduced.  Suppose ${\xi}(x,{\tau})$ is a transformed variable satisfying an imaginary time advection equation
\begin{equation}
\mathrm{i}\frac{{\partial}{\xi}}{{\partial}{\tau}}-\left[ \frac{1}{2}\frac{{\partial}}{{\partial}x}r(x,{\tau})^2\right] \frac{{\partial}{\xi}}{{\partial}x}=0.
\label{eq:ITADVEC}
\end{equation}
On the characteristic curve of Eq.~(\ref{eq:ITADVEC}), the variable coefficient NLSE in Eq.~(\ref{eq:VCNLSE}) is transformed into 
\begin{equation}
\frac{{\partial}{\varphi}}{{\partial}{\tau}}+\left(r({\xi},{\tau})\frac{{\partial}{\xi}}{{\partial}x}\frac{{\partial}}{{\partial}{\xi}}\right)^2{\varphi}+|{\varphi}|^2{\varphi}=0.
\label{eq:VCNLSE_II}
\end{equation}
Introducing the variable transformation as
\begin{equation}
{\eta}=\int_{}^{{\xi}}\frac{1}{r({\xi}',{\tau})}\left( \frac{{\partial}{\xi}'}{{\partial}x}\right)^{-1}\mathrm{d}{\xi}'
\label{eq:XItoETA}
\end{equation}
for Eq.~(\ref{eq:VCNLSE_II}) yields the NLSE with respect to ${\eta}$ and ${\tau}$ as
\begin{equation}
\frac{{\partial}{\varphi}}{{\partial}{\tau}}+\frac{{\partial}^2{\varphi}}{{\partial}{\eta}^2}+|{\varphi}|^2{\varphi}=0.
\label{eq:NLSE}
\end{equation}
Since the solutions of the NLSE have been presented by various methods, the inverse variable transformations with the solutions of Eq.~(\ref{eq:NLSE}) provide the corresponding solutions of the variable coefficient CGLE in Eq.~(\ref{eq:VCGLE}).
\subsection{Solvable condition of the imaginary time advection equation}
\label{subsec:2-2}
If the imaginary time advection equation in Eq.~(\ref{eq:ITADVEC}) is solved, the corresponding characteristic curve is obtained.  However, that is quite a difficult task, because the velocity term depends on both $x$ and $t$.  Hereafter, a solvable condition of the imaginary time advection equation is introduced, which enables derivation of the a closed form of the transformed variables.  \par
Suppose $p(x,t)$ and $q(x,t)$ are given by
\begin{eqnarray}
&&p(x,t)=X_p(x)T_p(t),
\label{eq:PSEP} \\
&&q(x,t)=X_q(x)T_q(t),
\label{eq:QSEP}
\end{eqnarray}
where $X_l(x)$ and $T_l(t)$ $(l=p,q)$ are $x$- and $t$-dependent functions, respectively.  With Eqs.~(\ref{eq:TFTAU}), (\ref{eq:PSEP}) and (\ref{eq:QSEP}), the imaginary time advection equation in Eq.~(\ref{eq:ITADVEC}) is rewritten as
\begin{equation}
\frac{\mathrm{i}}{T_p(t)}\frac{{\partial}{\xi}}{{\partial}t}-\frac{1}{2}X_q(x)\frac{{\partial}}{{\partial}x}\left(\frac{X_p(x)}{X_q(x)}\right)\frac{{\partial}{\xi}}{{\partial}x}=0.
\label{eq:SEPITADV}
\end{equation}
By the method of characteristics~\cite{Farlow1971}, the corresponding characteristic equations are derived as
\begin{eqnarray}
&&\frac{\mathrm{d}t}{\mathrm{d}s}=-\frac{\mathrm{i}}{T_p(t)},
\label{eq:CCT} \\
&&\frac{\mathrm{d}x}{\mathrm{d}s}=-\frac{1}{2}X_q(x) \frac{{\partial}}{{\partial}x}\left(\frac{X_p(x)}{X_q(x)}\right),
\label{eq:CCX}
\end{eqnarray}
where $s$ is an auxiliary variable of the characteristic curve.  From Eqs.~(\ref{eq:CCT}) and (\ref{eq:CCX}), the characteristic curve is derived as
\begin{equation}
{\kappa}=-\int_{}^{x}\frac{2}{X_q(x')}\left[ \frac{{\partial}}{{\partial}x}\left(\frac{X_p(x')}{X_q(x')}\right) \right]^{-1}\mathrm{d}x'+\mathrm{i}\int_{}^{t}T_p(t')\mathrm{d}t',
\label{eq:CC}
\end{equation}
and thus the solution of the imaginary time advection equation is obtained as
\begin{equation}
{\xi}(x,t)={\xi}_0({\kappa}),
\label{eq:SOLADV}
\end{equation}
where ${\xi}_0({\cdot})$ is an appropriate initial function of Eq.~(\ref{eq:ITADVEC}).  In addition, if $X_q(x)$ is constant, the transformed variable ${\eta}$ is evaluated simply as
\begin{equation}
{\eta}=\int_{}^{x}\frac{\mathrm{d}x'}{r(x',{\tau})}
\label{eq:ETAX}
\end{equation}
by the inversion function theorem.  Under these solvable conditions, the specific forms of transformed variables can be calculated.
\section{Review of exact solutions of the nonlinear Scr{\"o}dinger equation}
\label{sec:3}
In order to apply the proposed method in the previous section, the exact solutions of the NLSE are reviewed here.  The NLSE can be solved analytically by means of the Hirota's bilinear method.  The $D$-operator for differentiable functions $f(z)$ and $g(z)$ is defined by~\cite{Hirota1971}
\begin{equation}
D_z(f{\cdot}g)=\left[ \frac{{\partial}f(z)}{{\partial}z}g(z')-f(z)\frac{{\partial}g(z')}{{\partial}z'} \right]_{z'=z}. 
\label{eq:DefDOPR}
\end{equation}
The complex valued field function ${\varphi}({\eta},{\tau})$ is assumed to be the rational of real $F({\eta},{\tau})$ and complex $G({\eta},{\tau})$ functions as
\begin{equation}
{\varphi}({\eta},{\tau})=\frac{G({\eta},{\tau})}{F({\eta},{\tau})}.
\label{eq:GbyF}
\end{equation}
The following relations between partial derivatives and $D$-operators
\begin{eqnarray}
&&\frac{{\partial}{\varphi}}{{\partial}{\tau}}=\frac{D_{\tau}(G{\cdot}F)}{F^2},
\label{eq:DTAU} \\
&&\frac{{\partial}^2{\varphi}}{{\partial}{\eta}^2}=\frac{D_{\eta}^2(G{\cdot}F)}{F^2}-\frac{D_{\eta}^2(F{\cdot}F)}{GF}
\label{eq:DDETA}
\end{eqnarray}
for Eq.~(\ref{eq:NLSE}) yield the bilinear forms with respect to $F$ and $G$ as
\begin{eqnarray}
&&D_{\tau}(G{\cdot}F)+D_{\eta}^2(G{\cdot}F)={\lambda}GF,
\label{eq:BLF_GF} \\
&&D_{\eta}^2(F{\cdot}F)-|G|^2={\lambda}F^2,
\label{eq:BLF_FF}
\end{eqnarray}
where an arbitrary function ${\lambda}({\eta},{\tau})$ is introduced to incorporate the effect of boundary conditions in the bilinear forms in Eqs.~(\ref{eq:BLF_GF}) and (\ref{eq:BLF_FF}).  The special solutions of the NLSE are obtained from algebraic procedures for the bilinear forms.  \par
Suppose $F$ and $G$ are expanded with respect to an infinitesimal parameter ${\epsilon}$ as
\begin{eqnarray}
&&F=1+{\epsilon}^2F_2+{\epsilon}^4F_4+{\cdots}, 
\label{eq:EXPF} \\
&&G={\epsilon}G_1+{\epsilon}^3G_3+{\cdots}.
\label{eq:EXPG}
\end{eqnarray}
Substituting Eqs.~(\ref{eq:EXPF}) and (\ref{eq:EXPG}) into Eqs.~(\ref{eq:BLF_GF}) and (\ref{eq:BLF_FF}) generates an ${\epsilon}$-hierarchy of the bilinear forms.  Although the perturbation expansions in Eqs.~(\ref{eq:EXPF}) and  (\ref{eq:EXPG}) are infinite series, the ${\epsilon}$-hierarchy of the bilinear forms are truncated at a finite order.  In particular, the one soliton solution is immediately obtained under ${\lambda}=0$ as follows:
\begin{equation}
{\varphi}({\eta},{\tau})=A\mathrm{sech}(K{\eta}-V{\tau})\mathrm{e}^{\mathrm{i}{\theta}_0},
\label{eq:SOLITON}
\end{equation}
where $A$, $K$, $V$, and ${\theta}_0$ are constant parameters.  Since the NLSE describes the propagating envelope of a carrier wave, the soliton solution in Eq.~(\ref{eq:SOLITON}) is called an envelope soliton solution.  \par
The NLSE has a rational function solution~\cite{Peregrine1983}.  Suppose $F$ and $G$ in Eq.~(\ref{eq:GbyF}) are polynomial functions of $x$ and $t$, so their coefficients are determined sequentially by direct substitution.  Depending on the highest order of the polynomials, the rational function solutions of the NLSE exhibit different forms.  The possible lowest order polynomials of $F$ and $G$ yield the following rational function solution
\begin{equation}
{\varphi}({\eta},{\tau})=\left[1-\frac{4(1+2\mathrm{i}{\tau})}{1+4{\eta}^2+4{\tau}^2} \right]\mathrm{e}^{\mathrm{i}{\tau}}.
\label{eq:PSOLITON}
\end{equation}
The spatiotemporally localized soliton in Eq.~(\ref{eq:PSOLITON}) is known as the Peregrine soliton, which is used as a mathematical model of rogue waves and freak waves~\cite{Akhmediev2009a,Akhmediev2009b}.  \par
In addition, a periodic wave motion is derived from the NLSE~\cite{Akhmediev1987,Akhmediev1993}.  To obtain an oscillating solution an ansatz is introduced as follows:
\begin{equation}
{\varphi}({\eta},{\tau})=\left[{\rho}({\eta},{\tau})+{\sigma}({\tau})\right]\mathrm{e}^{\mathrm{i}{\theta}({\tau})},
\label{eq:ABANSATZ}
\end{equation}
where ${\rho}({\eta},{\tau})$ is a complex valued function, ${\sigma}({\tau})$ and ${\theta}({\tau})$ are real valued functions.  Substituting the ansatz in Eq.~(\ref{eq:ABANSATZ}) into the NLSE in Eq.~(\ref{eq:NLSE}) provides the set of differential equations with respect to ${\rho}$, ${\sigma}$ and ${\theta}$.  Through a cumbersome calculation under appropriate assumption of integrals, a periodic solution is obtained as
\begin{equation}
{\varphi}({\eta},{\tau})=\frac{{\cos}(\sqrt{2}{\eta})+\mathrm{i}\sqrt{2}{\sinh}{\tau}}{{\cos}(\sqrt{2}{\eta})-\sqrt{2}{\cosh}{\tau}}\mathrm{e}^{\mathrm{i}{\tau}}.
\label{eq:AKBREATHER}
\end{equation}
This periodic solution shows breathing of a localized wave, whereby it is called the Akhmediev breather.
\section{Example of variable transformations}
\label{sec:4}
Without loss of generality, an example of the variable transformations is demonstrated.  In this case, all of the coefficient functions in Eq.~(\ref{eq:VCGLE}) are given by elementary functions as follows:
\begin{eqnarray}
&&p(x,t) = (1+x^2)(1+t), 
\label{eq:VCP} \\
&&q(x,t) = \frac{1}{1+t},
\label{eq:VCQ} \\
&&{\gamma}(t) = \frac{1}{1+t},
\label{eq:VCGAMMA} \\
&&{\omega}(t)=1+{\cos}t.
\label{eq:VCOMEGA}
\end{eqnarray}
Although these coefficient functions do not originate from real physical systems, spatiotemporal modulations are often modelled by elementary functions.  In terms of the above coefficient functions, transformed functions and variables are derived as
\begin{eqnarray}
&&{\Gamma}(t) = {\log}(1+t),
\label{eq:TFGAMMA} \\
&&{\Omega}(t) = t+{\sin}t
\label{eq:TFOMEGA}
\end{eqnarray}
and
\begin{eqnarray}
&&{\xi}(x,t) = -\frac{1}{2}{\log}|x|+\mathrm{i}\left( t+\frac{1}{2}t^2\right) ,
\label{eq:EXAMXI} \\
&&{\eta}(x) = \frac{1}{2}{\log}|x|,
\label{eq:EXAMETA} \\
&&{\tau}(t) = t+\frac{1}{2}t^2.
\label{eq:EXAMTAU} 
\end{eqnarray}
In this case, ${\eta}$ depends only on $x$ because $T_p(t)=T_q(t)\mathrm{e}^{2{\Gamma}(t)}$ is satisfied.  By substituting ${\eta}(x)$ and ${\tau}(t)$ into the solutions of the NLSE presented in the previous section, the corresponding solutions of the variable coefficient CGLE can be obtained.  \par
This example is rather simple in order to demonstrate how to obtain the solution of the variable coefficient CGLE.  Of course, the proposed method can be utilized for other coefficient functions with numerical computations, beyond this case in which the variable transformations are completed analytically.  
\section{Related physical models}
\label{sec:5}
In the previous section an example of the transformed variables was demonstrated.  The coefficient functions were selected merely for simple calculations.  In this section, coefficient functions related to real physical systems have been proposed in the area of plasma, fluid, and optical systems. \par
Plasma systems are governed by electro-magnetic fluid equations.  As a slowly varying amplitude equation, the NLSE is derived from the original equations.  To incorporate inhomogeneity of media into slow dynamics, a variable coefficient NLSE has been proposed~\cite{Gao2001}. The proposed variable coefficient NLSE with reversion of independent variables is of the form:
 \begin{equation}
\mathrm{i}\frac{{\partial}{\psi}}{{\partial}t} + p(t)\frac{{\partial}^2{\psi}}{{\partial}x^2} + q(t)|{\psi}|^2{\psi} = 0,
\label{eq:EXAPLSM}
\end{equation}
where $p(t)$ and $q(t)$ are real valued functions.  The same model was used to describe chirp signal emissions in optical fibers~\cite{CHOW2008}.  \par
Related to Eq.~(\ref{eq:EXAPLSM}), a variable coefficient CGLE has been introduced as a model of Taylor instabilities or electron-beam plasma waves as follows~\cite{Liu2009}:
\begin{equation}
\mathrm{i}\frac{{\partial}{\psi}}{{\partial}t} + p\frac{{\partial}^2{\psi}}{{\partial}x^2} + q(t)|{\psi}|^2{\psi} = ({\omega} + \mathrm{i}{\gamma}){\psi},
\label{eq:EXAFLD}
\end{equation}
where $q(t)$ is a real valued function, and $p$, ${\gamma}$ and ${\theta}$ are constant real parameters.  By the transformations in Eqs.~(\ref{eq:PSItoPHI}), (\ref{eq:GAMMA}) and (\ref{eq:OMEGA}), it is confirmed that Eqs.~(\ref{eq:EXAPLSM}) and (\ref{eq:EXAFLD}) are congruent.  \par
In the inhomogeneous optical fiber, the dynamics of the optical pulse propagation is described by the following variable coefficient CGLE:
\begin{equation}
\mathrm{i}\frac{{\partial}{\psi}}{{\partial}t} - \left[\frac{1}{2}p_r(t)+\mathrm{i}p_i(t)\right] \frac{{\partial}^2{\psi}}{{\partial}x^2} + \left[q_r(t)-\mathrm{i}q_i(t)\right]|{\psi}|^2{\psi} =\mathrm{i}{\gamma}(t){\psi},
\label{eq:EXAOPT}
\end{equation}
where $p_r(t)$, $p_i(t)$, $q_r(t)$, $q_i(t)$, and ${\gamma}(t)$ are real valued functions.  Note that in the original system $x$ and $t$ in Eq.~(\ref{eq:EXAOPT}) denote time and space, respectively.  \par
Since all the physical models in this section are special cases of the variable coefficient CGLE in Eq.~(\ref{eq:VCGLE}), the proposed analytical method can be applied to obtain the solutions of those models.
\section{Conclusion}
\label{sec:6}
In this paper, the analytical method for solving the variable coefficient CGLE is presented.  On the characteristic curve, which makes the imaginary time advection equation invariant, the variable coefficient CGLE is transformed into the NLSE with respected to the transformed variables.  Also, the solvable condition for the imaginary time advection equation is introduced in order to obtain the characteristic curve analytically.  An example of the coefficient functions is demonstrated for yielding a closed form of the transformed variables without loss of generality.  It is confirmed that the related physical models are special cases of the variable coefficient CGLE considered in this paper.  \par
On the other hand, a space-dependent coefficient function for the linear term was not considered.  That means, as a model of the Bose-Einstein condensation, Eq.~(\ref{eq:VCGLE}) is not applicable to describe the spatiotemporal dynamics of macroscopic wave functions.  Incorporating such effect into the variable coefficient CGLE is a future work.

%
%



\begin{thebibliography}{}
%
%
\bibitem{Cross1993}
Cross, M.C., Hohenberg, P.C.: Pattern formation outside of equilibrium. Rev. Mod. Phys. {\bf 65}, 851–1112 (1993)
\bibitem{Aranson2002}
Aranson, I.S., Kramer, L.: The world of the complex Ginzburg–Landau equation. Rev. Mod. Phys. {\bf 74}, 99–143 (2002)
\bibitem{Daniels2004}
Daniels, K.E., Beck, C., Bodenschatz, E.: Defect turbulence and generalized statistical mechanics. Physica D {\bf 193}, 208–217 (2004)
\bibitem{Akhmediev2005}
Akhmediev, N., Ankiewicz, A.: Dissipatve solitons. Springer, Berlin (2005)
\bibitem{Beta2006}
Beta, C., Mikhailov, A. S., Rotermund, H. H., Ertl, G.: Defect-mediated turbulence in a catalytic surface reaction. Europhys. Lett. {\bf 75}, 868–874 (2006)
\bibitem{Bekki2012}
Bekki, N., Harada, Y., Kanai, H.: Bekki-Nozaki hole in traveling excited waves on human cardiac interventricular septum. J. Phys. Soc. Jpn. {\bf 81}, 073801 (2012)
\bibitem{Kuramoto1984}
Kuramoto, Y.: Chemical oscillations, waves, and turbulence. Springer, Berlin (1984) 
\bibitem{Ablowitz1974}
Ablowitz, M. J., Kaup, D. J., Newell, A. C., Segur. H.: Inverse scattering transform-Fourier analysis for nonlinear problems. Stud. Appl. Phys. {\bf 53}, 249-315 (1974)
\bibitem{Hirota1971}
Hirota, R.: Exact solution of the Korteweg-de Vries equation for multiple collisions of solitons. Phys. Rev. Lett. {\bf 27}, 1192-1194 (1971)
\bibitem{Bekki1984}
Nozaki, K., Bekki, N.: Exact solutions of the generalized Ginzburg-Landau equation. J. Phys. Soc. Jap. {\bf 53}, 1581-1582 (1984)
\bibitem{Pethick2001}
Pethick, C. J., Smith, H.: Bose–Einstein condensation in dilute gases. Cambridge University Press, Cambridge (2001)
\bibitem{Li2007}
Li, Z. D., et. al.: Hirota method for the nonlinear Schr{\"o}dinger equation with an arbitrary linear time-dependent potential. Ann. Phys. {\bf 322}, 2545-2553 (2007)
\bibitem{Bronski1}
Bronski, J. C., et. al.: Bose-Einstein condensates in standing waves: the cubic nonlinear Schr{\"o}dinger equation with a periodic potential. Phys. Rev. Lett. {\bf 86}, 1402-1405 (2001)
\bibitem{Mallory2013}
Mallory, K., Van Gorder, R. A.: Stationary solutions for the 1 + 1 nonlinear Schr{\"o}dinger equation modeling repulsive Bose-Einstein condensates in small potentials. Phys. Rev. E {\bf 88}, 013205 (2013) 
\bibitem{Gao2001}
Gao, Y. T.: Variable-coefficient unstable nonlinear Schr{\"o}dinger equation modeling electron beam plasma: Auto-B{\"a}cklund transformation, soliton-typed and other analytical solutions. Phys. Plasmas. {\bf 8}, 67-73 (2001)
\bibitem{Liu2009}
Liu, J. G., Li, Y. Z., Tian., B.: Soliton-like solutions for the modified variable-coefficient Ginzburg–Landau equation. Comm. Nonlinear Sci. Num. Sim. {\bf 14}, 1214–1226 (2009)
\bibitem{Fang2006}
Fang, F., Xiao, Y.: Stability of chirped bright and dark soliton-like solutions of the cubic complex Ginzburg–Landau equation with variable coefficients. Opt. Comm. {\bf 268}, 305–310 (2006)
\bibitem{Wong2015}
Wong, P., et. al.: Dromion-like structures and stability analysis in the variable coefficients complex Ginzburg–Landau equation. Ann. Phys. {\bf 360}, 341–348 (2015)
\bibitem{CHOW2008}
CHOW, K. W., et. al.: Transmission and stability of solitary pulses in complex Ginzburg–Landau equations with variable coefficients. J. Phys. Soc. Jpn. {\bf 77}, 054001 (2008)
\bibitem{Liu2017}
Liu, W., et. al.: Analytic solutions for the generalized complex Ginzburg–Landau equation in fiber lasers. Nonlinear Dyn. {\bf 89}, 2933–2939 (2017)
\bibitem{Farlow1971}
Farlow, S. J.: Partial differential equations for scientists and engineers. John Wiley \& Sons, New York (1971)
\bibitem{Peregrine1983}
Peregrine, D. H.: Water waves, nonlinear Schr{\"o}dinger equations and their solutions. Austral. Math. Soc. Ser. B {\bf 25}, 16-43 (1983)
\bibitem{Akhmediev2009a}
Akhmediev, N., Ankiewicz, A., Taki, M.: Waves that appear from nowhere and disappear without a trace. Phys. Lett. A {\bf 373}, 675-678 (2009)
\bibitem{Akhmediev2009b}
Akhmediev, N., Soto-Crespo, J. M., Ankiewicz, A.: Extreme waves that appear from nowhere: On the nature of rogue waves. Phys. Lett. A {\bf 373}, 2137-2145 (2009)
\bibitem{Akhmediev1987}
Akhmediev, N. N., Eleonskii, V. M., Kulagin, N. E.: Exact first-order solutions of the nonlinear Schr{\"o}dinger equation. Theoret. Math. Phys. {\bf 72}, 809–818 (1987)
\bibitem{Akhmediev1993}
Akhmediev, N., Ankiewicz, A.: First-order exact solutions of the nonlinear Schr{\"o}dinger equation in the normal-dispersion regime. Phys. Rev. A. {\bf 47}, 3213-3221 (1993)

\end{thebibliography}


\end{document}